\documentclass[12pt]{article}

\usepackage{times}
\usepackage{times, xcolor, graphicx, graphics, cite}

\newcounter{lastnote}

\begin{document} 


\baselineskip24pt

\begin{center}
	\LARGE{Deformation Behaviour of Ion-Irradiated FeCr: \\A Nanoindentation Study} 
	\bigskip
	
	
	\large
	{Kay Song$^{1}$\footnote[1]{kay.song@eng.ox.ac.uk}, Hongbing Yu$^{2}$, Phani Karamched$^{3}$, Kenichiro Mizohata$^{4}$, \\
		David E. J. Armstrong$^{3}$, Felix Hofmann $^{1}$\footnote[2]{felix.hofmann@eng.ox.ac.uk}}\\
	
		\bigskip
		\small{$^{1}$ Department of Engineering Science, University of Oxford, Parks Road, Oxford, OX1 3PJ, UK} \\
		\small{$^{2}$ Canadian Nuclear Laboratories, Chalk River, ON K0J 1J0, Canada} \\
		\small{$^{3}$ Department of Materials, University of Oxford, Parks Road, Oxford OX1 3PH, UK} \\
		\small{$^{4}$ University of Helsinki, P.O. Box 64, 00560 Helsinki, Finland}

\end{center}

\begin{abstract}
Understanding the mechanisms of plasticity in structural steels is essential for the operation of next-generation fusion reactors. Elemental composition, particularly the amount of Cr present, and irradiation can have separate and synergistic effects on the mechanical properties of ferritic/martensitic steels. The study of ion-irradiated FeCr alloys is useful for gaining a mechanistic understanding of irradiation damage in steels. Previous studies of ion-irradiated FeCr did not clearly distinguish between the nucleation of dislocations to initiate plasticity, and their propagation through the material as plasticity progresses. 

In this study, Fe3Cr, Fe5Cr, and Fe10Cr were irradiated with 20 MeV Fe$^{3+}$ ions at room temperature to nominal doses of 0.01 dpa and 0.1 dpa. Nanoindentation was carried out with Berkovich and spherical indenter tips to study the nucleation of dislocations and their subsequent propagation. The presence of irradiation-induced defects reduced the theoretical shear stress and barrier for dislocation nucleation. The presence of Cr further enhanced this effect due to increased retention of irradiation defects. However, this combined effect is still small compared to dislocation nucleation from pre-existing sources such as Frank-Read sources and grain boundaries. The yield strength, an indicator of dislocation mobility, of FeCr increased with irradiation damage and Cr. The increased retention of irradiation defects due to the presence of Cr also further increased the yield strength. Reduced work hardening capacity was also observed following irradiation. The synergistic effects of Cr and irradiation damage in FeCr appear to be more important for the propagation of dislocations, rather than their nucleation.


\end{abstract}



\section{Introduction}

Reduced activation ferritic/martensitic (RAFM) steels are prime candidate materials for structural components of next-generation nuclear fusion and fission reactors \cite{Baluc2004}. They are favoured over austenitic steels for their resistance to irradiation swelling and helium embrittlement, and generally good thermomechanical properties \cite{Ehrlich2001, Tavassoli2002}. For the safe and efficient operation of reactors, the impact of the reactor environment, mainly the effect of neutron irradiation, on the behaviour of the reactor materials must be well-understood. For the structural components of the reactor, it is particularly important to characterise the mechanical properties and the onset of plasticity and deformation.

The study of FeCr binary alloy materials is useful for gaining mechanistic insight into the effects of irradiation on RAFM steels as they reduce the microstructural complexity associated with numerous minor alloying elements in industrial steels \cite{Garner2000, Matijasevic2006, Lambrecht2011}.  Ion-irradiation is a useful surrogate for simulating damage to materials from neutron irradiation. It allows the accumulation of large damage doses, in a controlled environment, over a short span of time without inducing transmutation in the samples \cite{Was2014, Harrison2019}. This eliminates evolution in the chemical composition of the samples and allows for comparison to simulations.  The drawback to using ion-irradiation instead of neutrons is that the damage layer produced is only a few microns thick, which necessitates the use of specialised techniques to characterise the material properties at such scales.

Nanoindentation has proved to be invaluable for providing mechanical information of ion-irradiation damaged materials as it can extract mechanical information from the thin irradiated layers \cite{Weaver2018, Heintze2011, Bushby2012, Armstrong2013, Kasada2014, Gao2019}. It has been used to investigate mechanical properties including yield strengths, deformation behaviour and elastic properties \cite{Armstrong2015, Kiener2012, Pohl2019, Xiao2020}.

Previous nanoindentation studies on irradiated FeCr, with different Cr\%, largely focused on quantifying the irradiation-induced hardening behaviour \cite{Hardie2013a, Hardie2013, Heintze2011, Song2020}. At room temperature, irradiation hardening has been observed in FeCr, even at damage levels as low as 0.01 displacements-per-atom (dpa) \cite{Song2020}. The amount of hardening has been found to increase with Cr content \cite{Heintze2011, Song2020}. There exists some experimental work on the deformation behaviour of irradiated FeCr, particularly after the onset of yield, with strain softening commonly observed \cite{Bushby2012, Armstrong2015, Hardie2012, Hardie2015, Hardie2016}. An unanswered question from the current literature is the behaviour just before and at the onset of pop-in when the dominant effects are from dislocation nucleation and multiplication. The effect of irradiation on the mechanisms responsible for the initiation of plasticity in FeCr is also currently unclear. Both the initiation and progression of plasticity are crucial to predicting the structural integrity of the steels in operation.

A second focus missing in existing studies is the synergistic effect of Cr and irradiation damage on the initiation and progression of plasticity in FeCr. Existing studies \cite{Bushby2012, Armstrong2015, Hardie2012} have focused only on one composition of Cr\% or one level of damage in each study, making comparisons of the effect of Cr and damage level difficult, as samples from different studies vary in processing history, irradiation condition and data analysis protocol. From our previous study of irradiated FeCr \cite{Song2020}, we found that enhanced defect retention from the increasing presence of Cr caused greater changes in the hardness, thermal diffusivity and lattice strain. This present study follows on from these previous results to focus on the synergistic effects of Cr and irradiation dose level on the deformation behaviour of irradiated FeCr.


Here we systematically examine the deformation behaviour of FeCr with nanoindentation as a function of Cr content (3, 5, and 10\%), and damage level (unimplanted, 0.01, and 0.1 dpa). Irradiation and mechanical characterisation were carried out at room temperature, providing a different picture of microstructural evolution to more common studies with irradiation at 300$^{\circ}$C. Nanoindentation reveals the dose- and composition-dependent behaviour of dislocation nucleation and propagation in the irradiation-damaged regions of the materials to provide insights on the early stage deformation behaviour of irradiated FeCr. These findings are discussed in light of previous FeCr deformation studies and ab-initio predictions of deformation behaviour. This information is crucial for the design and performance prediction of components in future reactors.

\section{Materials and Methods}
\subsection{Materials and ion-implantation}
This investigation considers the same samples as \cite{Song2020}, which contains full details of their preparation and history. Briefly, polycrystalline samples of FeCr with 3, 5, and 10\% Cr content (here referred to as Fe3Cr, Fe5Cr, and Fe10Cr, respectively) were manufactured under the European Fusion Development Agreement (EFDA) programme by induction melting and a series of forging and recrystallisation heat treatments \cite{Coze2007}. The samples were mechanically ground with SiC paper then polished with diamond suspension and colloidal silica (0.04 $\mu$m). The final surface treatment performed was electropolishing with 5\% perchloric acid in ethanol at -40$^{\circ}$C using a voltage of 28 V for 3-4 minutes.

Ion-implantation was performed with 20 MeV Fe$^{3+}$ ions at room temperature using the 5 MV tandem accelerator at the Helsinki Accelerator Laboratory. For each sample composition, samples were damaged to 2 nominal damage doses: 0.01 dpa and 0.1 dpa (by averaging over the first 2 $\mu$m of damage), as calculated using the Quick K-P model in the SRIM code (20 MeV Fe ions on a Fe target with 25 eV displacement energy at normal incidence) \cite{Ziegler2010}. The calculated damage layer extends to 3 $\mu$m below the sample surface (Figure \ref{fig:stress_profile}). Previous measurements using X-ray diffraction confirmed that the depth-dependent changes in the levels of strain in this layer match the predictions from SRIM \cite{Song2020}. One sample of each FeCr composition was set aside as an unimplanted reference.

\subsection{Nanoindentation procedures}
Nanoindentation experiments were performed using an MTS Nano Indenter XP. A Berkovich tip, and two spherical tips of nominal radius 1 $\mu$m and 5 $\mu$m respectively were used. Continuous stiffness measurements (CSM) were conducted with a frequency of 45 Hz and a harmonic amplitude of 2 nm. The indenter, for all tip shapes, was loaded at a constant value of $({\mathrm{d}P}/{\mathrm{d}t})/P$ = 0.05 s$^{-1}$. Grains of near $\langle 100\rangle$, $\langle 110\rangle$ and $\langle 111\rangle$ out-of-plane orientations were first identified by electron backscatter diffraction, then chosen for indentation. At least 10 indents were performed on each sample.



\subsection{Nanoindentation analysis}
\begin{figure}
	\includegraphics[width=\textwidth]{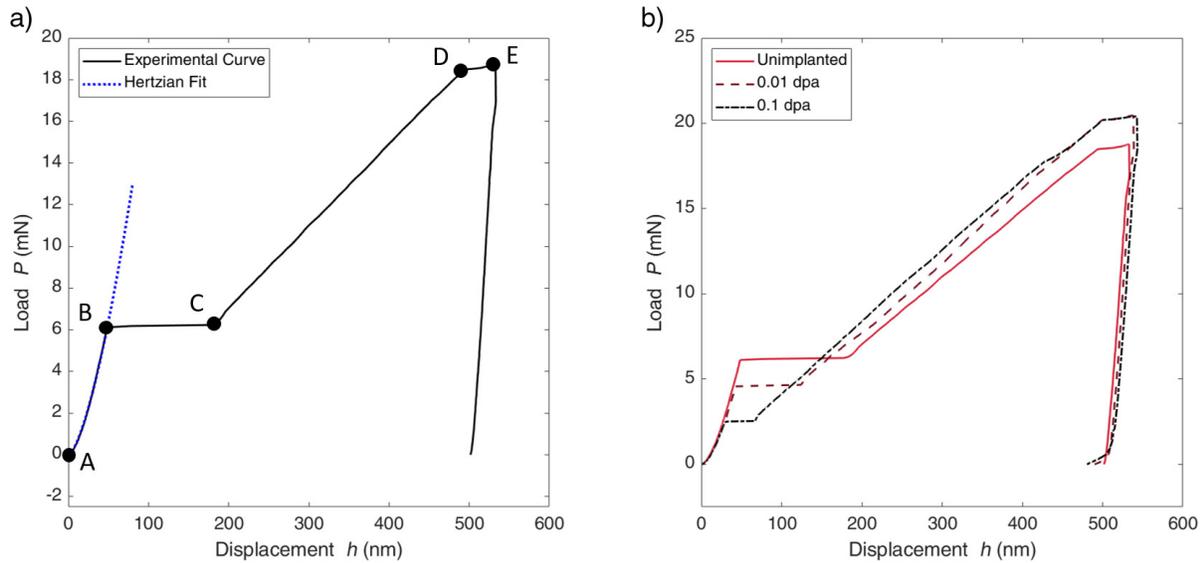}
	\centering
	\caption{The load-displacement curve of the indentation of Fe3Cr with a nominally 5 $\mu$m radius spherical tip. a) The indentation curve of unirradiated Fe3Cr with key segments labelled: elastic deformation described by Hertzian mechanics (A $\rightarrow$ B), pop-in (B $\rightarrow$ C), elastic-plastic deformation (C $\rightarrow$ D) and creep (D $\rightarrow$ E). b) The effect of irradiation on increasing the pop-in loads for Fe3Cr.} 
	\label{fig:p-hcurve}
\end{figure}

A typical load vs. displacement curve measured in this experiment is shown in Figure \ref{fig:p-hcurve}. The first section of the curve between points A and B is the elastic response which can be described as the contact between a sphere (the indenter tip) and an elastic half-space (the sample surface) \cite{Johnson1987}:
\begin{equation}\label{eqn:Hertz}
P = \frac{4}{3} E_{r}R^{\frac{1}{2}}h_{e}{}^{\frac{3}{2}}
\end{equation}
where $P$ is the applied load, $R$ is the radius of the sphere, $h_{e}$ is the elastic indentation depth. $E_{r}$ is the reduced modulus given by:
\begin{equation}
\frac{1}{E_{r}} = \frac{1-\nu_{i}{}^{2}}{E_{i}} + \frac{1-\nu_{s}{}^{2}}{E_{s}}
\end{equation}
where $E$ is the elastic modulus and $\nu$ is the Poisson's ratio, with subscripts $i$ and $s$ referring to the indenter and the sample respectively. Here we used $E_{i}$ = 1000 GPa, $\nu_{i}$ = 0.07 \cite{Fischer-Cripps2011} and $\nu_{s}$ = 0.29 \cite{Ledbetter1973}. From our previous study \cite{Song2020}, $E_{s}$ was found to be 205 GPa for Fe3Cr and Fe5Cr, and 215 GPa for Fe10Cr.

By fitting the elastic portion of the load-displacement curve (between points A to B in Figure \ref{fig:p-hcurve}) to Equation \ref{eqn:Hertz}, the effective radius ($R_{eff}$) for the Berkovich tip was found to be 120 $\pm$ 31 nm. The spherical tips were found to have effective radii of 800 $\pm$ 120 nm and 5.12 $\pm$ 0.13 $\mu$m. Note, the manufacturer reported 1.11 $\pm$ 0.04 $\mu$m and 4.83 $\pm$ 0.24 $\mu$m, respectively, for the spherical indenter tips.

A `pop-in' occurs between points B and C (Figure \ref{fig:p-hcurve}a), and the pop-in load (at point B), $P_{pop}$ was identified for all samples. Elastic-plastic deformation (between points C and D) occurs following a pop-in event as dislocations have been nucleated in the material. Sample creep can also be observed between points D and E when the indenter is held at the maximum load for 10 seconds.

The use of tips with different effective radii created differently sized stress zones underneath the surface. The principal shear stress ($\tau$) directly ahead of the indenter tip in the sample before the occurrence of pop-ins was computed using Hertzian contact mechanics \cite{Johnson1987}:
\begin{equation}\label{eqn:pss}
	\tau = \frac{1}{2} p_{0} \left| (1+\nu) \left(1-\left(\frac{z}{a}\right)\arctan\left(\frac{a}{z}\right)\right) - \frac{3}{2} \left(\frac{1}{1+z^{2}/a^{2}}\right)\right|
\end{equation}
where $z$ is the coordinate into the material normal to the surface and $a$ is the contact radius, further discussed below in Equation \ref{eqn:contactradius}. $p_{0}$ is given by \cite{Johnson1987}:
\begin{equation} \label{eqn: maxss}
p_{0} = \left(\frac{6P_{pop}E_{r}{}^{2}}{\pi^{3}R^{2}}\right)^{\frac{1}{3}}.
\end{equation}

The maximum shear stress experienced by the sample at the onset of pop-ins is given by \cite{Johnson1987}:
\begin{equation}\label{eqn:maxss1}
\tau_{max} = 0.31 p_{0}.
\end{equation}

\subsubsection*{Indentation stress-strain curves} \label{sec:iss}
Starting from the elastic contact described in Equation \ref{eqn:Hertz}, and defining contact radius $a = \sqrt{R h_{e}}$, we can transform Equation \ref{eqn:Hertz} into a linear relationship during elastic contact \cite{Kalidindi2008, Pathak2015}. Indentation stress, $\sigma_{ind}$, and strain, $\epsilon_{ind}$, in the elastic regime, are defined as:

\begin{equation}
\sigma_{ind} = \frac{P}{\pi a^{2}}; \;\;\; \epsilon_{ind} = \frac{4}{3\pi} \frac{h_{e}}{a}
\end{equation}

For elastic-plastic indentation, $h_{t}$, the total indentation depth, is used instead of $h_{e}$, the elastic indentation depth. In this picture, it is equivalent to idealising the indentation zone as a cylindrical region, of radius $a$ and height $\frac{3\pi}{4}a$, being compressed by $h_{t}$ \cite{Kalidindi2008, Pathak2015}.

The contact radius $a$ is calculated from the measured stiffness, $S = dP/dh_{e}$, at each point in the loading curve:
\begin{equation}\label{eqn:contactradius}
a = \frac{S}{2E_{r}}
\end{equation}

Corrections for the point of effectively zero load and displacement were also made according to methods described in \cite{Kalidindi2008}. By considering the initial elastic segment ($A \rightarrow B$) in the loading curve and using Hertz's theory, the stiffness can be written as:
\begin{equation}\label{eqn:corr}
S = \frac{3P}{2h_{e}} = \frac{3(\tilde{P}-P^{\star})}{2(\tilde{h_{e}}-h^{\star})}
\end{equation}

where $\tilde{P}$ and $\tilde{h_{e}}$ are the measured load and displacement signals. $P^{\star}$ and $h^{\star}$ are the values of load and displacement at the point of effective initial contact. By rearranging Equation \ref{eqn:corr}, plotting $\tilde{P}-\frac{2}{3}S\tilde{h_{e}}$ against $S$ produced a linear relationship with a slope of $-\frac{2}{3}h^{\star}$ and a \mbox{y-intercept} of $P^{\star}$. The values of $P^{\star}$ and $h^{\star}$ were then found by linear regression.

\subsection{High-resolution electron backscatter diffraction (HR-EBSD)}
Cross-correlation analysis of high angular resolution EBSD/Kikuchi patterns can be used to extract lattice rotations with respect to a reference point in an EBSD map. These rotations can be used to calculate local lattice curvatures and then compute geometrically necessary dislocation (GND) densities on the surface of the material with a spatial resolution of $\sim$ 50 nanometres and a sensitivity above $10^{12}$ lines/m$^{2}$ \cite{Jiang2013, Wilkinson2006}. 

HR-EBSD measurements were performed on a Zeiss Merlin SEM (20 kV, 15 nA), with a Bruker eFlash detector system. The sample was tilted by 70$^{\circ}$ with respect to the electron beam, and a working distance of 18 mm was used. The measurements used a step size of 170 nm to ensure sufficient spatial coverage over the area around an indent while maintaining reasonable measurement times. Diffraction patterns were recorded at a resolution of 800 $\times$ 600 pixels. The cross-correlation analysis was carried out using a MATLAB-based code, originally written by Britton and Wilkinson \cite{Britton2011, Britton2012}.

\subsection{Electron channelling contrast imaging (ECCI)}
Electron channelling contrast strongly relies on the backscattered electron intensity signal while the electrons travel in a sample volume. The presence of lattice distortions, such as dislocations, can alter the diffraction condition and thus allow defects to be imaged under specific conditions \cite{Wilkinson1997, Zaefferer2014}. This method is able to detect both GND and statistically stored dislocations (SSD) in the material. ECCI measurements were performed on a Zeiss Crossbeam 540 SEM (30 kV, 10 nA), with a four-quadrant backscattered electron detector at nearly normal incidence to the electron beam direction.

\section{Results}

\subsection{Pop-in shear stress analysis}

\begin{figure}
	\includegraphics[width=0.6\textwidth]{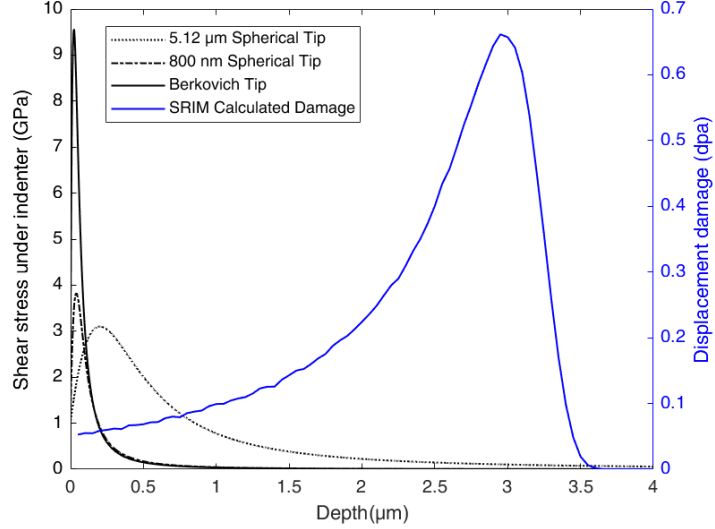}
	\centering
	\caption{The principle shear stress directly ahead of the indenter tip at the onset of pop-in for Fe10Cr irradiated to 0.1 dpa (black traces). The size of the highly stressed zone increases with tip size but are all still contained in the irradiated region (blue trace).} 
	\label{fig:stress_profile}
\end{figure}

Figure \ref{fig:stress_profile} shows the principal shear stress as a function of depth for different indenter tip radii, calculated using Equations \ref{eqn:pss} and \ref{eqn: maxss} for the loads recorded at the onset of pop-in events. It is plotted for Fe10Cr irradiated to 0.1 dpa, but the features are representative of all the samples in this study. As the tip radius increases, the stressed zone extends deeper into the material. Even for the largest tip, 5.12 $\mu$m spherical, the main stressed zone is well-contained within the ion-irradiated layer. 

The maximum shear stress under the indenter ($\tau_{MUI}$) at pop-in also decreases with increasing tip size. This has been observed previously in nickel \cite{Shim2008}, tungsten \cite{Beake2018} and molybdenum \cite{Morris2011}. This size-dependent phenomenon originates from the initiation of plasticity following either the nucleation of dislocations or the activation of existing dislocations (e.g. Frank-Read sources) in the material. The former mechanism generally requires a much higher applied stress than the latter. The volume of the highly stressed zone scales with the indenter radius, as shown in Figure \ref{fig:stress_profile}. The probability of encountering a pre-existing dislocation is thus higher in the stressed zone under a larger indenter, compared to applying the same load to a smaller indenter. Plasticity initiates once the applied stress is large enough to mobilise pre-existing defects \cite{Shim2008}. For a smaller indenter, it is more likely that plasticity initiates only when the applied stress is greater than the ideal strength of the material so that new shear dislocation loops can be nucleated.

\begin{figure}
	\includegraphics[width=0.6\textwidth]{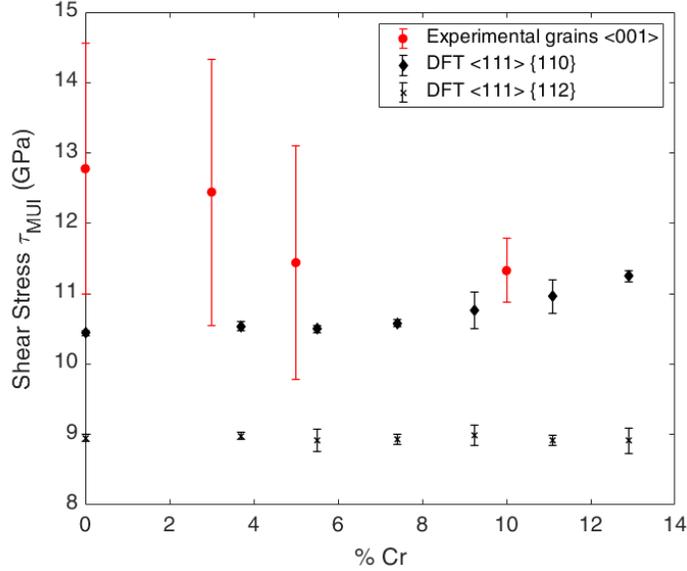}
	\centering
	\caption{The maximum shear stress ($\tau_{MUI}$) under the Berkovich indenter tip ($R_{eff}$ = 120 nm) at the onset of pop-in for unimplanted FeCr with different levels of Cr\% for grains of near $\langle 100\rangle$ out-of-plane orientation (red circles), where the primary slip system is $\langle 111\rangle \{110\}$. The error bars show one standard deviation of the measurements. Density functional theory (DFT) calculations (black diamonds and crosses) \cite{Casillas-Trujillo2017} are included for comparison.} 
	\label{fig:SS_cr}
\end{figure}

The $\tau_{MUI}$ at the onset of pop-in during loading with a Berkovich tip for the unimplanted materials is calculated using Equation \ref{eqn: maxss} and \ref{eqn:maxss1} (Figure \ref{fig:SS_cr}). In the small, relatively defect-free volume probed by the sharp Berkovich tip, $\tau_{MUI}$ at the onset of pop-in should approach the theoretical shear strength. This has been observed before in several other material systems \cite{Gouldstone2000, Dub2017, Ahn2012, Shim2008}. Experimental values shown in Figure \ref{fig:SS_cr} are only from indenting grains with close to $\langle 100 \rangle$ out-of-plane orientation, and the main corresponding slip system is $\langle 111\rangle \{110\}$. In our study, the experimental $\tau_{MUI}$ is very close to the theoretical shear strength calculated from density functional theory \cite{Casillas-Trujillo2017}. One possible source of discrepancy could be variable surface roughness between samples. The elastic modulus value of the material, which was used in the calculations from Equation \ref{eqn: maxss}, was measured from nanoindentation and could have been affected by the presence of pile-up \cite{Beck2017}.

Our experimental measurements of $\tau_{MUI}$ at pop-in also shows a monotonic decrease with increasing Cr content. This suggests that the presence of Cr causes dislocations to nucleate at a lower applied stress. The effect of alloying elements on the theoretical strength of materials has been previously investigated in theoretical studies, for example in titanium \cite{Song1999} and silicon \cite{Huang1994} and the effect depends strongly on the material matrix and solute. For BCC metals at ambient temperatures, plasticity is governed by kink-pair nucleation mechanism and kink migration \cite{Shinzato2019}. An increase of solute concentration in BCC iron has been found to reduce the nucleation barrier of kink-pairs \cite{Shinzato2019, Ghafarollahi2020}, which is consistent with the dependence of $\tau_{MUI}$ on Cr concentration in FeCr observed here.

\begin{figure}
	\includegraphics[width=0.6\textwidth]{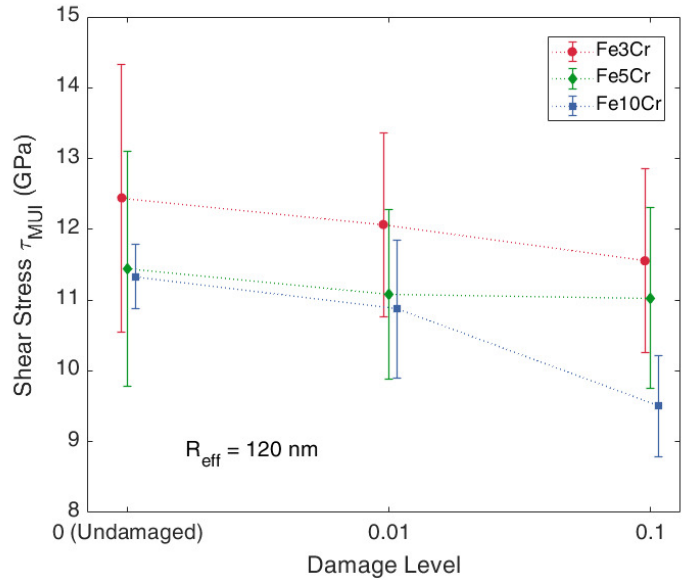}
	\centering
	\caption{The maximum shear stress under the Berkovich indenter tip ($\tau_{MUI}$) at the onset of pop-in for Fe3Cr (red circle), Fe5Cr (green diamond) and Fe10Cr (blue square) as a function of damage. The error bars show one standard deviation of the measurements.} 
	\label{fig:SS_berk}
\end{figure}

The effect of ion-irradiation on pop-in load, and thus $\tau_{MUI}$ at the onset of pop-in, can be seen in Figure \ref{fig:SS_berk}. The pop-in loads were not found to vary systematically between grains of near $\langle 100\rangle$, $\langle 110\rangle$ and $\langle 111\rangle$ out-of-plane orientations, and from here on, results from all orientations are considered together. 

The decrease in $\tau_{MUI}$ with irradiation damage can be justified by considering irradiation-induced defects in the material as sites for heterogeneous dislocation nucleation. It has been shown before, in Ni-irradiated Mo, that ion-irradiation-induced defects serve as local instabilities, which decrease the shear stress required for dislocation nucleation \cite{Jin2018}. This differs from other pop-in studies which investigated materials damaged by mechanical straining, where the main form of heterogeneous dislocation nucleation is from Frank-Read sources with much lower activation barriers than irradiation-induced defects \cite{Bei2016}. It also appears that the effect of Cr content (Figure \ref{fig:SS_cr}) on the change in $\tau_{MUI}$ is greater than the effect of solely increasing irradiation damage (Figure \ref{fig:SS_berk}). 

The synergistic effects of Cr\% and irradiation damage can be seen in the greater decrease of $\tau_{MUI}$ for Fe10Cr compared to Fe3Cr and Fe5Cr following irradiation. This can be attributed to greater retention of irradiation-induced defects in the presence of Cr atoms \cite{Prokhodtseva2013, Hernandez-Mayoral2008}. The irradiation-induced change of other material properties such as hardening, decrease in thermal diffusivity and increase in irradiation strain has also been found to be greater for higher Cr\% in irradiated FeCr \cite{Song2020}.

\begin{figure}
	\includegraphics[width=\textwidth]{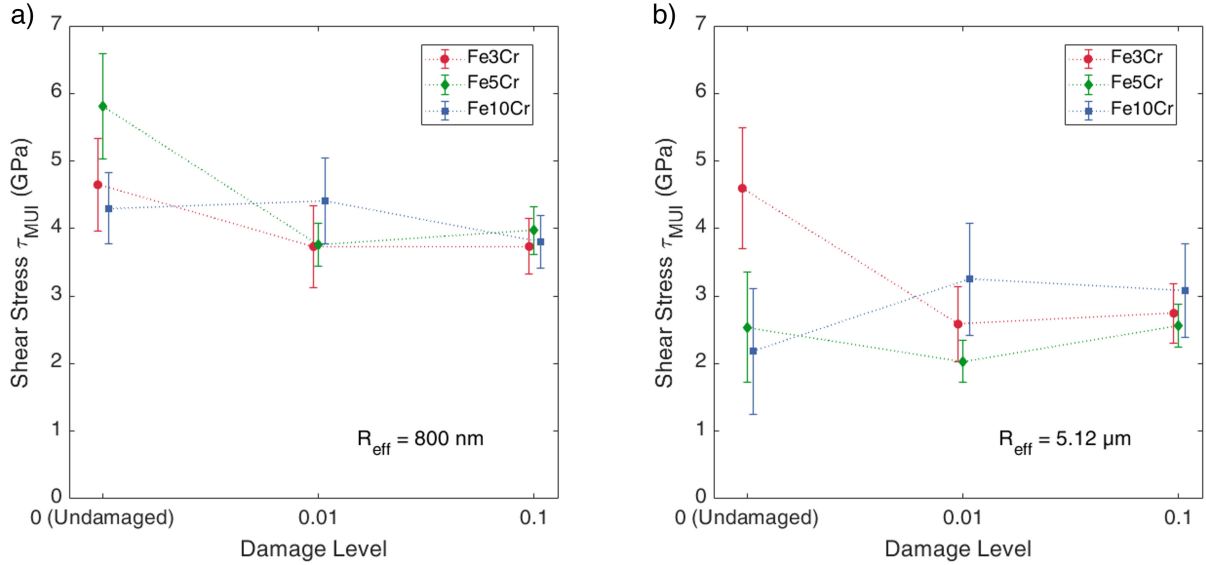}
	\centering
	\caption{The maximum shear stress under the indenter tip ($\tau_{MUI}$) at the onset of pop-in for Fe3Cr (red circle), Fe5Cr (green diamond) and Fe10Cr (blue square) as a function of damage for spherical indenters with effective radii a) 800 nm and b) 5.12 $\mu$m. The error bars show one standard deviation of the measurements.} 
	\label{fig:SS_spher}
\end{figure}

For indentation performed with larger spherical tips, with $R_{eff}$ = 800 nm and 5.12 $\mu$m, pop-ins were still very noticeable in both the reference (undamaged) and irradiated samples. $\tau_{MUI}$ at the onset of pop-in for these spherical tips are shown in Figure \ref{fig:SS_spher}. For Fe3Cr and Fe5Cr, indenting with the larger spherical tips still reveals a reduction in $\tau_{MUI}$ with increasing damage levels, similar to the observations from the Berkovich indentation pop-ins in Figure \ref{fig:SS_berk}.

For spherical tips, the $\tau_{MUI}$ measured are all significantly lower than those for the Berkovich tip, as discussed above. This drop in $\tau_{MUI}$ for increased tip radius is much larger than the reduction in $\tau_{MUI}$ following irradiation (seen in both Figures \ref{fig:SS_berk} and \ref{fig:SS_spher}). This suggests that the heterogeneous dislocation nucleation sources probed during indentation with larger tips have lower activation barriers than dislocation nucleation from irradiation-induced defects. 

For Fe10Cr, $\tau_{MUI}$ at the onset of pop-in is seen to increase following irradiation damage when probed by a spherical tip with an effective radius of 5.12 $\mu$m (Figure \ref{fig:SS_spher}b), which is the opposite trend to the dependence of $\tau_{MUI}$ on irradiation damage for all other samples. This may point to more complex behaviour of dislocation nucleation following irradiation from heterogeneous sources with higher activation barriers. The other possible explanation is that the unimplanted Fe10Cr material possesses an unusually high level of pre-existing dislocations, which causes a particularly low level of $\tau_{MUI}$ required for pop-ins and initiation of plasticity. 

\begin{figure}
	\centering
	\includegraphics[width=\textwidth]{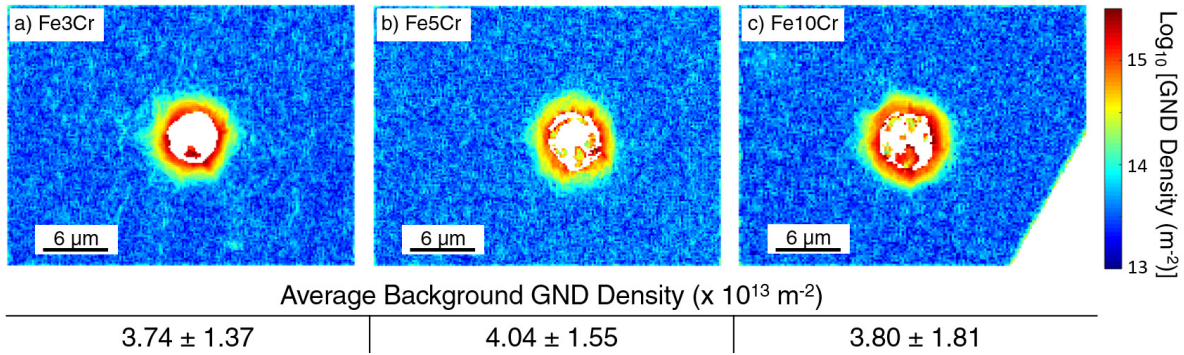}
	\caption{GND density distribution around spherical indents ($R_{eff}$ = 5.12 $\mu$m) in grains of near $\langle 100 \rangle$ out-of-plane orientation on unimplanted a) Fe3Cr, b) Fe5Cr, and c) Fe10Cr as measured by HR-EBSD. The average and standard deviation of the GND density are displayed below the corresponding maps.}
	\label{fig:hrebsd}
\end{figure}

HR-EBSD measurements (Figure \ref{fig:hrebsd}) show the GND densities in the vicinity of an indent on each of the FeCr unimplanted surfaces. The average background GND density is stated below the corresponding maps. No differences between samples of different Cr concentrations could be found after accounting for the measurement uncertainties.

\begin{figure}
	\centering
	\includegraphics[width=\textwidth]{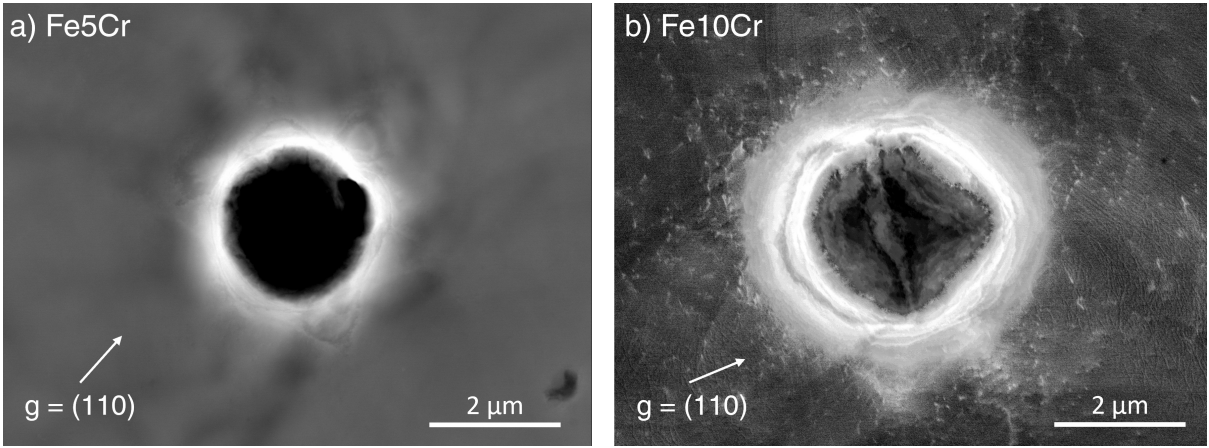}
	\caption{ECCI maps of regions around spherical indents ($R_{eff}$ = 5.12 $\mu$m) in grains of near $\langle 111 \rangle$ out-of-plane orientation on unimplanted a) Fe5Cr, and b) Fe10Cr. Some dislocations can be observed around the indent in Fe10Cr but the contrast does not appear to be strong enough to observe dislocations in Fe5Cr.} 
	\label{fig:ecci}
\end{figure}

ECCI was also used to image and measure the dislocation densities in the material, as it can probe both GNDs and SSDs. The ECCI maps of indents made in unimplanted Fe5Cr and Fe10Cr are shown in Figure \ref{fig:ecci}. Fe10Cr appears to exhibit slightly higher dislocation density than Fe5Cr in the region $\sim$1 to 2 microns away from the indent. Unfortunately, the backscatter contrast signal was not strong enough in the less deformed and undeformed regions to determine a background dislocation density. From the images, the pre-existing dislocation density in unimplanted Fe5Cr and Fe10Cr seem to be quite low, though a quantitative estimate seems difficult.



\subsection{Indentation stress-strain curves}

\begin{figure}
	\centering
	\includegraphics[width=\textwidth]{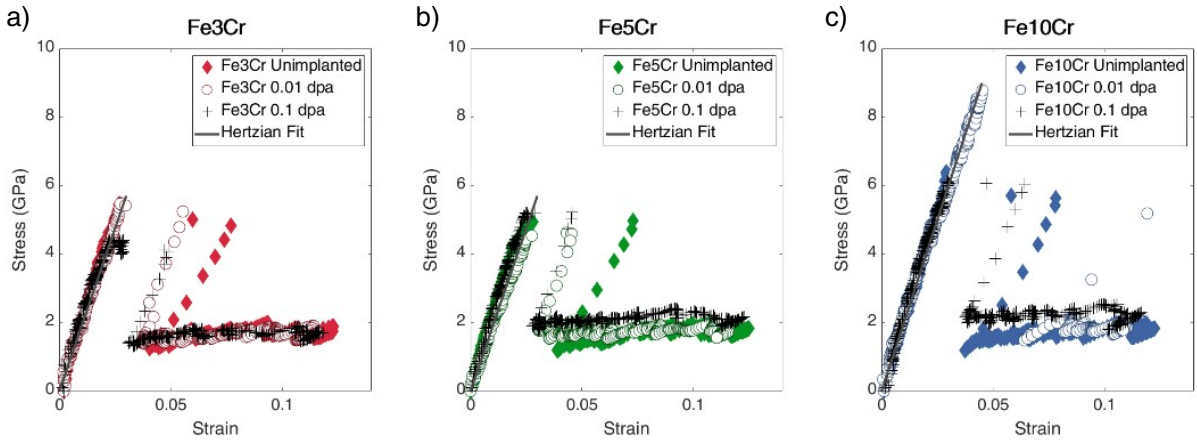}
	\caption{The indentation stress-strain responses of a) Fe3Cr, b) Fe5Cr, and c) Fe10Cr irradiated to different damage levels. The elastic portions of the curves are well-fitted to the prediction from Hertzian contact mechanics (Equation \ref{eqn:Hertz}) with the gradient equal to the reduced modulus of each sample. } 
	\label{fig:iss_total}
\end{figure}

Using the protocols described in Section \ref{sec:iss} \cite{Pathak2017, Patel2016}, indentation stress-strain (ISS) curves were calculated for the indents made with the 5.12 $\mu$m spherical tip (Figure \ref{fig:iss_total}). The effects of pop-ins are clearly seen, with large elastic stresses in the ISS response before the flow portion proceeds at much lower stress levels. The stresses seen in the flow portions of the ISS response depend on damage level for Fe5Cr and Fe10Cr. Previous studies have shown that pop-ins do not affect the plastic behaviour in the ISS response as this depends on the mobility of dislocations rather than their nucleation \cite{Pathak2009, Pathak2015}. 

However, the presence of the pop-ins presents some difficulties in the analysis of the ISS response as they obstruct direct determination of the yield point and plastic behaviour immediately following yield. The method described in \cite{Zhu2008} was used to infer the yield point as the intersection of the elastic Hertzian response with a straight line fitted through the flow portion of the ISS curve. The yield strength is then defined using a 0.2\% strain offset protocol. This involves shifting the ISS curve by 0.2\% and finding the intersection between the elastic portion of the shifted curve and the straight line fit of the flow portion of the original ISS curve. This protocol is commonly used for defining yield strength in uniaxial tensile testing but has also been used in many studies to identify yield strength in nanoindentation measurements \cite{Weaver2016a, Weaver2016, Patel2016}. 

\begin{figure}
	\centering
	\includegraphics[width=0.6\textwidth]{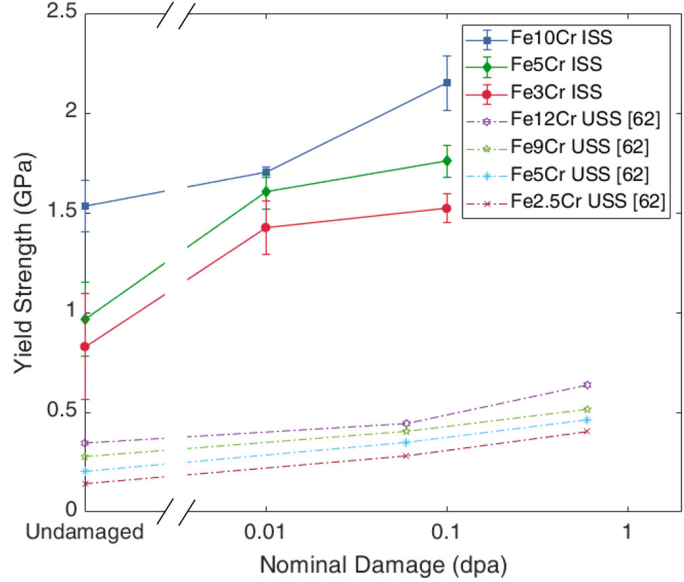}
	\caption{The yield strengths measured from the ISS responses of the FeCr material as a function of damage in this study, using $R_{eff} = $ 5.12 $\mu$m (markers with solid lines). The error bars show one standard deviation of the measurements. A comparison with yield strength measured from USS responses \cite{Matijasevic2008} (markers with dashed lines) is shown. } 
	\label{fig:ys}
\end{figure}

The yield strength calculated from ISS response increases monotonically with Cr\% for all damage levels (Figure \ref{fig:ys}). The origins of this, for tests all performed at the same temperature, is solid solution hardening, which increases with the square root of the solute concentration \cite{Dieter1961, Xiao2016}. The yield strength also increases monotonically with irradiation level for all FeCr materials measured, which has been previously observed in other irradiated materials \cite{Pathak2017, Matijasevic2008}. 


The rate of yield strength increase between unimplanted and 0.01 dpa of damage is lower for Fe10Cr compared to Fe3Cr and Fe5Cr, which would be consistent with a slightly higher pre-existing dislocation density in the Fe10Cr material, as discussed above. The higher rate of yield strength increase for Fe10Cr, compared to Fe3Cr and Fe5Cr, between 0.01 dpa and 0.1 dpa of damage, suggests increased retention of defects by the presence of Cr from a similar mechanism to irradiation hardening observed previously \cite{Matijasevic2008, Song2020}.

These trends also agree with results from uniaxial stress-strain (USS) responses of neutron-irradiated FeCr \cite{Matijasevic2008} (also Figure \ref{fig:ys}). The discrepancy between the absolute values of the two studies can be attributed to several factors. The material on which USS response was measured was irradiated by neutrons at around 300$^{\circ}$C which could have partially annealed the material, compared to the materials in this study which was ion-irradiated at room temperature. 

Previous studies to correlate ISS and USS responses have suggested a linear relationship between the yield strengths obtained by each method:
\begin{equation}
Y_{ISS} = mY_{USS}
\end{equation}

where $m$ is between 2 to 3, depending on the protocols used to define the ISS response and the anisotropy of the material studied \cite{Tabor1948, Herbert2001, Basu2006, Donohue2012, Weaver2018}. This factor is mainly to account for the lateral constraint of materials in the ISS response, which is absent in the USS response.
For the definitions of indentation stress and strain used in this study, a value of $m$ between 1.9 to 2.2 has been found to obtain good agreement with USS responses for a range of metals \cite{Weaver2016a, Patel2016}. This brings the absolute values of the yield strengths from the USS response closer to the values measured from our study. Furthermore, the method of extrapolating the yield point may slightly over-estimate the yield strength as the gradient of the ISS curve near the yield point is often higher than the gradient at larger strains. 

Another point to consider is the dependence of yield stress on the indenter size $R$. This indentation size effect has the same origin as increasing hardness values measured at decreasing depth for pyramidal indenters due to the increased density of geometrically necessary dislocations \cite{Nix1998, Swadener2002, Durst2008}. In some metals, this has been found to scale by $R^{-1/3}$ \cite{Gerberich2002, Spary2006} but for well-annealed metals, this effect has been found to be negligible \cite{Vachhani2015, Weaver2016a, Weaver2016b}. 


\begin{figure}
	\centering
	\includegraphics[width=0.6\textwidth]{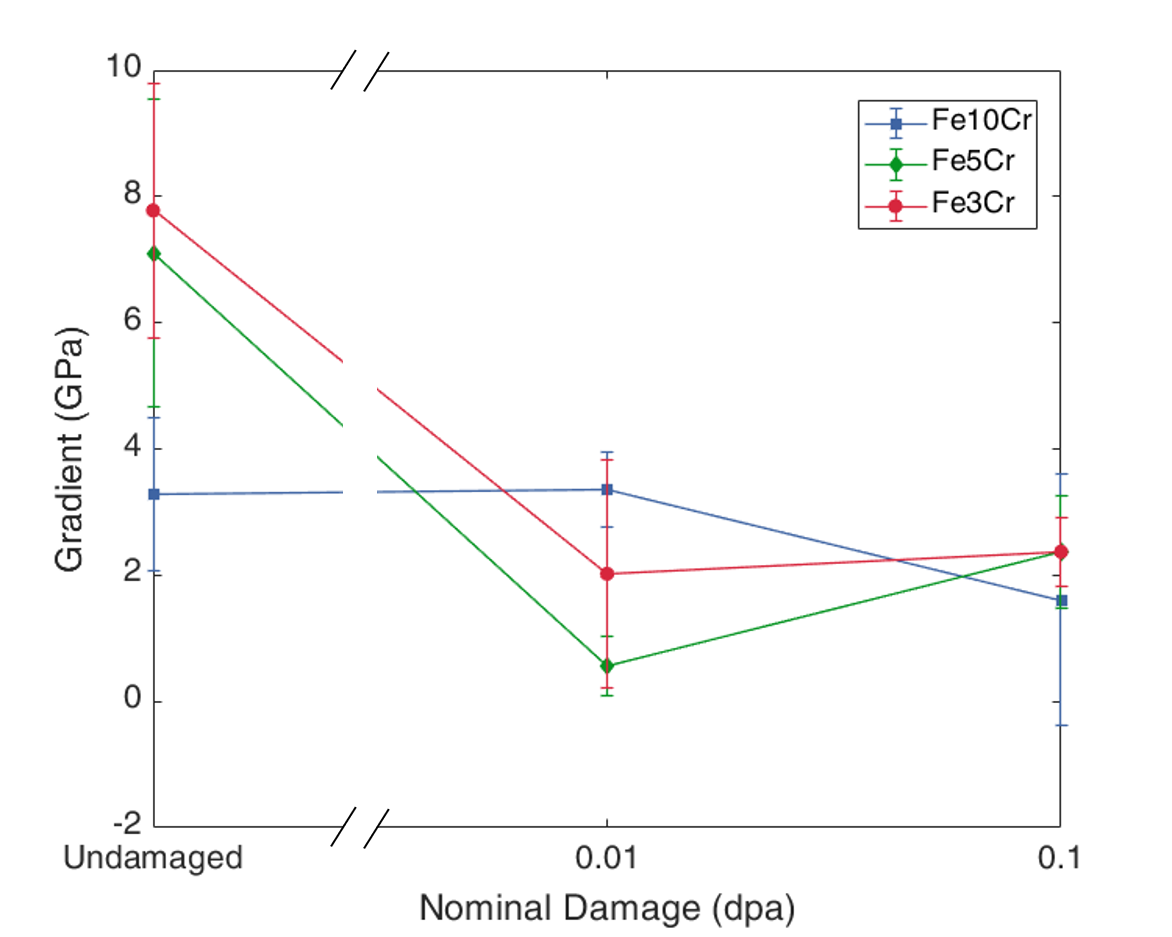}
	\caption{The gradient of the flow portion from the ISS response of FeCr materials as a function of damage level. The error bars show one standard deviation of the measurements.} 
	\label{fig:grad}
\end{figure}

The shape of the ISS curve in the flow portion gives insight into material deformation behaviour, similar to the work hardening coefficient \cite{Armstrong2015}. The gradient of a linear fit to the flow portion for the different FeCr materials as a function of damage is shown in Figure \ref{fig:grad}. The flow portion gradient appears to decrease after irradiation for all Cr content. From previous bulk tensile tests, a decrease in work hardening has also been observed in the uniaxial stress-strain responses of FeCr following neutron-irradiation \cite{Matijasevic2008}. Analysis of the ISS response and TEM imaging of the plastic zone under indents in ion-irradiated Fe12Cr revealed that strain softening occurs beyond the yield point from the annihilation of irradiation defects by reactions with glissile dislocations \cite{Hardie2012, Bushby2012}.

The yield behaviour observed is similar to that previously described for helium-implanted tungsten \cite{Das2019}. It exhibits initial irradiation hardening due to the interactions of the dislocations with the irradiation-induced defects. Our observations of increased yield strength in irradiated FeCr also shows a similar behaviour, with the formation of slip-steps in the pile-up topography around the indentation site to support this hypothesis. With increasing deformation, strain softening occurs in helium-implanted tungsten due to the removal of irradiation-induced helium defects by the glide dislocations. The reduced flow portion gradient in the ISS response of the irradiated FeCr suggests a similar mechanism may be occurring in the present FeCr samples \cite{Bacon2006}. 

\subsection{Characterising behaviour in the initial stages of deformation}
By combining the insights gathered from the nanoindentation pop-in trends and indentation stress-strain response studies, we can formulate a coherent picture of the early stages of deformation and plasticity in irradiated FeCr.

In a volume of FeCr material free of pre-existing defects, homogeneous dislocation nucleation occurs when the applied stress exceeds the theoretical strength of the material. The presence of Cr has been found to decrease the theoretical strength of FeCr, reducing the barrier for dislocation nucleation. Ion-irradiation causes the formation of defects. At low doses, they act as local instabilities in the lattice, reducing the shear stress required to nucleate dislocations. Though their effect on reducing the shear strength of the material is weak compared to that of Cr atoms, it is enhanced by the presence of Cr, which increases the retention of irradiation-induced defects \cite{Yao2008, Hernandez-Mayoral2008}. However, other pre-existing heterogeneous dislocation nucleation sources, such as Frank-Read sources and grain boundaries still play a more important role in nucleating dislocations in FeCr, as they require a lower stress for activation. 

Once the dislocations have nucleated, their relative mobility in different materials can be determined through the yield strength. The presence of Cr, even in unirradiated materials, reduces the mobility of dislocations, due to solid solution hardening. Following irradiation, the induced defects also act to hinder dislocation motion, further increasing the yield strength. Cr atoms again act to further enhance this by increasing the retention of irradiation-induced defects, causing a larger rate of increase of yield strength following irradiation for materials with high Cr\%. The increased retention of defects due to Cr also causes a greater decrease in the work-hardening capacity of the irradiated FeCr material due to the interactions between irradiation-induced defects and glide dislocations \cite{Das2019}.

It appears that the presence of Cr increasing the retention of irradiation-induced defects has a strong effect on the initial stages of deformation and plasticity in irradiated FeCr more through hindering dislocation motion than dislocation nucleation.

\section{Summary and Conclusion}
Fe3Cr, Fe5Cr, and Fe10Cr have been irradiated with 20 MeV Fe$^{3+}$ ions to nominal damage levels of 0.01 dpa and 0.1 dpa at room temperature. The effects of Cr content and irradiation dose on the initial stages of plasticity were studied using nanoindentation. We were able to separate the effects on initiation of plasticity and its early stages of progression. We conclude the following from our findings:
\begin{itemize}
	\item The presence of irradiation-induced defects reduces the theoretical strength of irradiated FeCr by reducing the barrier shear stress required for dislocation nucleation.
	\item The presence of Cr acts in a similar way but has a greater effect than solely that of irradiation-induced defects. A synergistic effect is found whereby the increased retention of irradiation-induced defects due to the presence of the Cr causes the largest reduction in the shear stress required to initiate plasticity.
	\item The effective barrier for dislocation nucleation is the lowest for pre-existing heterogeneous sources in the material, such as Frank-Read sources and grain boundaries, compared to nucleation from irradiation-defects or lattice instabilities caused by the presence of Cr.
	\item From the indentation stress-strain responses, the yield strength of the FeCr materials were found to increase with both Cr content and irradiation damage. The increased retention of irradiation-induced defects due to the presence of Cr further increased the yield strength. 
	\item These trends agree with uniaxial stress-strain responses of neutron-irradiated FeCr material. This demonstrates the usefulness of nanoindentation to characterise the bulk mechanical properties of irradiated materials.
	\item A decrease in work hardening is found following irradiation, which is more evident for samples with higher Cr content due to a greater population of defects, which interact with glide dislocations.
\end{itemize}

All data, raw and processed, as well as the processing and plotting scripts are available at: \textit{A link will be provided after the review process and before publication.}

\section{Acknowledgements}
The authors would like to thank Robert Scales, Yun Deng, and Anna Kareer from the Department of Materials, University of Oxford for their assistance with ECCI image processing, pop-in searching algorithms, and ISS calculations. The  authors  acknowledge  use  of  characterisation  facilities  within  the  David  Cockayne  Centre for  Electron  Microscopy,  Department  of  Materials,  University  of  Oxford,  alongside  financial support  provided  by  the  Henry  Royce  Institute (Grant  ref EP/R010145/1). KS acknowledges funding from the General Sir John Monash Foundation and the University of Oxford Department of Engineering Science. FH acknowledges funding from the European Research Council (ERC) under the European Union’s Horizon 2020 research and innovation programme (grant agreement No. 714697). DEJA acknowledges funding from EPSRC grant EP/P001645/1.
\bibliographystyle{unsrt}
\bibliography{ref}

\end{document}